\documentclass[twocolumn,secnumarabic,amssymb, nobibnotes, aps, prd]{revtex4-1}
\usepackage{graphicx}
\usepackage{color}
\usepackage{epsfig}

\begin{document}

\title{Spectral weight redistribution in (LaNiO$_3$)$_n$/(LaMnO$_3$)$_2$ superlattices from optical spectroscopy}

\author{P. Di Pietro$^1$, J. Hoffman$^2$, A. Bhattacharya$^2$, S. Lupi$^3$, A. Perucchi$^1$}

\affiliation{$^1$INSTM Udr Trieste-ST and Sincrotrone Trieste, Area Science Park, I-34012
Trieste, Italy} 
\affiliation{$^2$Materials Science Division, Argonne National Laboratory, Argonne, Illinois 60439} 
\affiliation{$^3$CNR-IOM  and Dipartimento di Fisica, Universit\`a di Roma
Sapienza,  P.le Aldo Moro 2, I-00185 Roma, Italy}

\date{\today}

\date{\today}

\begin{abstract}

We have studied the optical properties of four (LaNiO$_3$)$_n$/(LaMnO$_3$)$_2$ superlattices (SL) ($n$=2, 3, 4, 5) on SrTiO$_3$ substrates. We have measured the reflectivity at temperatures from 20 K to 400 K, and extracted the optical conductivity through a fitting procedure based on a Kramers-Kronig consistent Lorentz-Drude model.  With increasing LaNiO$_3$ thickness, the SLs undergo an insulator-to-metal transition (IMT) that is accompanied by the transfer of spectral weight from high to low frequency. The presence of a broad mid-infrared band, however, shows that the optical conductivity of the (LaNiO$_3$)$_n$/(LaMnO$_3$)$_2$ SLs is not a linear combination of the LaMnO$_3$ and LaNiO$_3$ conductivities. Our observations suggest that interfacial charge transfer leads to an IMT due to a change in valence at the Mn and Ni sites.

\end{abstract}

\maketitle

Recent progress in the growth of correlated oxide heterostructures has sparked interest in understanding and exploiting the novel electronic and magnetic properties that emerge at the interfaces between different material systems \cite{mannhart}. In transition metal oxides, electronic orbitals tend to overlap less than the $s$- and $p$- orbitals in semiconductors, resulting in strong electronic correlations \cite{imada}, higher values of the effective mass of charge carriers, and larger coupling to the polarized lattice. Correlation effects can be further enhanced at oxide interfaces, since the rupture of the periodicity of the ion lattice at the interface may reduce the electronic screening and thus increase the on-site Coulomb interaction. Interfacial charge redistribution in these systems results in the reconstruction of orbital and spin degrees of freedom (electronic reconstruction) producing correlated  two-dimensional (2D) states with novel electronic and magnetic behaviors \cite{bhatta}.

We address here (LaNiO$_3$)$_n$/(LaMnO$_3$)$_2$ superlattices \cite{hoffman13}, where the subscript denotes the layer thickness in terms of pseudocubic unit cells. X-ray spectroscopy of this system shows that the Mn oxidation state is converted from 3+ to 4+, while Ni is intermediate between 2+ and 3+ \cite{hoffman13}. A metal to insulator transition is observed as $n$ is decreased from 5 to 2. Charge transfer at the LNO/LMO interface therefore provides the opportunity to control the interplay between magnetism and charge transport in a Ni based oxide, as highlighted by several recent theoretical \cite{lee, dong} and experimental \cite{gibert, hoffman13} studies.

LaNiO$_3$ (LNO) is the only member of the perovskite rare-earth nickelate series that does not undergo a metal to insulator transition at low temperatures (T) \cite{torrance92}. Due to the low-spin 3$d^7$ configuration of Ni$^{3+}$, it has been proposed that the combination of LNO with other oxides in some superlattice (SL) structure may provide the possibility to mimic the CuO$_2$ planes of high-temperature superconductors \cite{chaloupka08,hansmann09}. LaMnO$_3$ (LMO) on the other hand, is an antiferromagnetic insulator, known for being the parent compound of colossal magnetoresistance manganites \cite{tokura99,tokura06}.

In this work we have measured the reflectivity of four (LNO)$_n$/(LMO)$_2$ SLs ($n$=2, 3, 4, 5) together with a LNO thin film, from 100 to 16000 cm$^{-1}$ at nearly normal incidence and between 20 and 400 K. The measurements were performed by means of a BRUKER 70v interferometer, using a gold mirror as a reference and various beamsplitters, detectors and thermal sources to cover the whole infrared range. The samples were grown by molecular beam epitaxy, as previously described in Ref. \cite{hoffman13}.  LNO and LMO thin films prepared under the same conditions as the superlattices were found to possess electronic and magnetic properties comparable to high-quality bulk single-crystal samples, suggesting the samples in this study are stoichiometric.  By varying the number of bilayers within each superlattice, the total thickness of the samples was kept constant at about 64 nm, as confirmed by x-ray reflectivity measurements. Our reflectivity measurements elucidate the mechanism behind the metal insulator transition with decreasing $n$. In particular, we observe a systematic shift in the spectral weight away from the Drude peak, and also from higher frequencies (near-infrared and above) into the mid-infrared, with the onset of insulating behavior as $n$ decreases from 5 to 2. Our observations are consistent with a localization mechanism that is driven by interfacial transfer of electrons from LMO into LNO.\\

In Fig. 1 we report the reflectivities of the four SLs, together with that of LNO, from 100 to 8000 cm$^{-1}$. Due to the small thickness of the SLs, the reflectivity is dominated by the STO substrate as shown by the gray line in Fig. 1a. Below 800 cm$^{-1}$, i.e., in the far-infrared region characterized by the large STO phonon band, the reflectivity decreases with increasing $n$. On the other hand, the opposite occurs in the mid-infrared (800-8000 cm$^{-1}$), where the reflectivity of the SL is enhanced with respect to the substrate. A similar effect was previously observed in the (LaMnO$_3$)/(SrMnO$_3$) series \cite{perucchi}, where an increase (decrease) in reflectivity in the mid- (far-) infrared range is attributed to higher conductivity properties. The reflectivity of the $n$=2 superlattice does not vary significantly with temperature. As shown in Fig. 1b, the temperature dependence of the $n$=3 compound is essentially confined to frequencies above 1000 cm$^{-1}$. As $n$ increases, the observed T-dependence progressively moves towards the far-infrared range, with the $n$=5 and pure LNO samples being independent of temperature above 2000 cm$^{-1}$. 

\begin{figure}
\includegraphics[width=9cm]{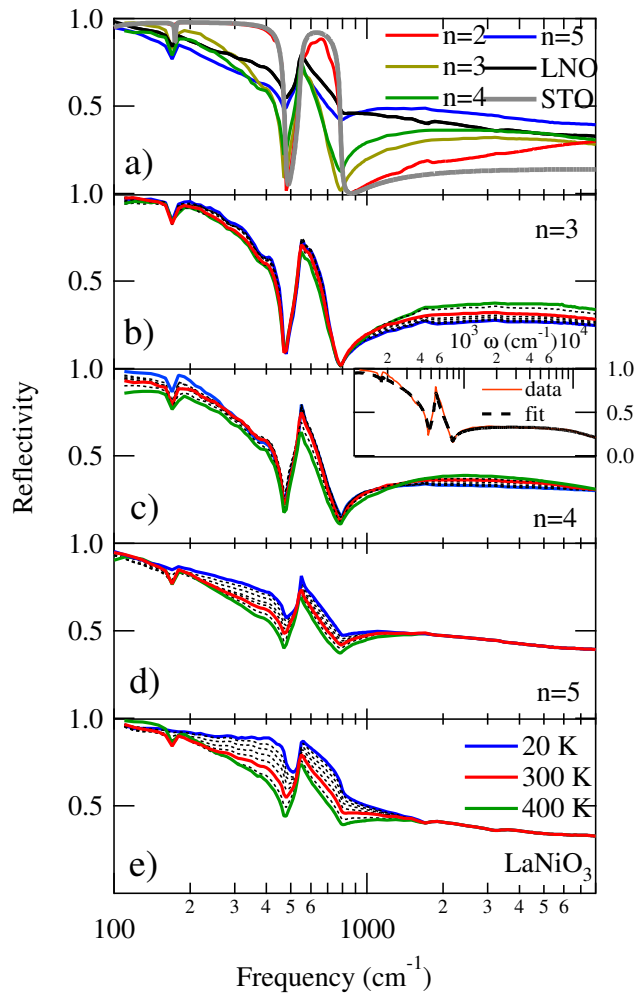}
\caption{Reflectivity of the LNO/LMO superlattices. a) Room temperature reflectivities of the $n$=2,3,4,5 samples together with the reflectivity of pure LNO and that of the STO substrate. 
b) Temperature dependence of the reflectivity of the $n$=3 compound. c-e) Same for the $n$=4,5 and for pure LNO. Dashed lines correspond to intermediate temperatures 100, 150, 200, 250, and 350 K. In the inset of panel c) the data at 20 K and the relative fit are reported in the whole range.}
\label{Fig.1}
\end{figure}

The optical conductivity of the LNO/LMO superlattices can be extracted through a fitting procedure based on the Kramers-Kronig consistent Lorentz-Drude model \cite{kuzmenko05}, by taking into account the finite thickness of the sample and the substrate's contribution to the reflectivity \cite{perucchi}. An overview of all the room temperature optical conductivities is provided in Fig. 2a, after subtracting the phonon modes of the SL. This helps to single out the electronic contribution to the optical conductivity.

For the $n$=2 compound, the optical conductivity smoothly tends to zero in the dc limit, thus indicating insulating behavior. However, the onset of a mid-infrared band is clearly present. We note that in pure LMO the first absorption band is centered at about 2.5 eV (20000 cm$^{-1}$), while no spectral weight is found in the mid-infrared region \cite{okimoto97}. At larger values of $n$, the optical conductivity increases in the whole infrared range, as a signature for the onset of metallization. For the $n$=5 compound, a clear Drude term is present, superimposed on an almost flat background infrared conductivity. Interestingly, the $n$=5 SL differs from pure LNO, which displays a sharper Drude peak and lower mid-infrared spectral weight.

In panels b to e in Fig. 2, we report the T-dependence of the optical conductivities in the various compounds. For low $n$, the T-dependence is more pronounced in the mid-infrared range, while with increasing $n$, the T-induced changes gradually shift towards the far-infrared. We observe a non-negligible increase of the mid-infrared spectral weight, with increasing temperature for the low $n$ SLs. For larger $n$, as well as for pure LNO, the Drude peak lowers and slightly broadens, with increasing temperature.

\begin{figure}
\includegraphics[width=9cm]{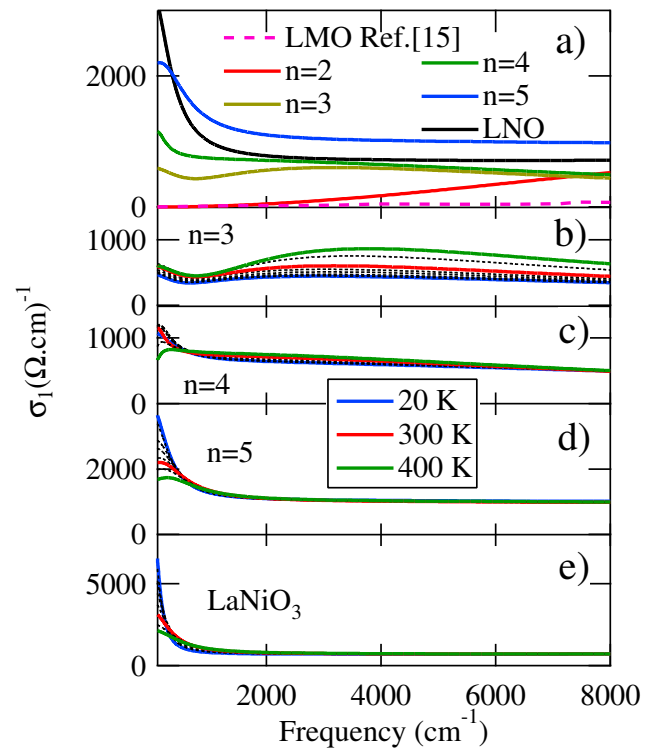}
\caption{Optical conductivity of the LNO/LMO superlattices as extracted from a Lorentz-Drude fitting. a) Room temperature optical conductivities of the $n$=2,3,4,5 samples and pure LNO and LMO. b) Temperature dependence of the optical conductivity of the $n$=3 compound. c-e) Same for the $n$=4,5 and for pure LNO. }
\label{Fig2}
\end{figure}

\begin{figure}
\includegraphics[width=9cm]{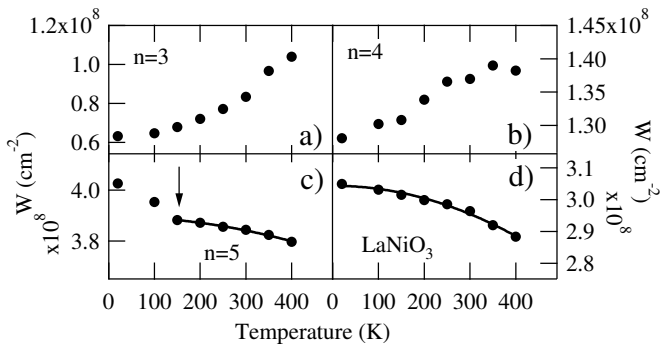}
\caption{Temperature dependence of the spectral weight $W(T)$ integrated up to $\omega_p$, for the $n$=3,4,5 and LNO compounds. For $n$=3 and $n$=4, $W(T)$  increases with increasing temperature, thus signaling the presence of incoherent excitations at low energies. For both $n$=5 and LNO the spectral weight can be fitted (continuous line) according to the quadratic relation $W(T)=W_0-BT^2$.}
\label{Fig2b}
\end{figure}

The LNO film displays metallic properties, with dc conductivity values reaching up to 10$^{4}$ ($\Omega$.cm)$^{-1}$, in good agreement with previously reported values of resistivity \cite{hoffman13}. A clear Drude peak coexists with a broad electronic background associated with interband transitions within the Ni $3d$ manifold \cite{basov}. The plasma frequency ($\omega_p$), calculated as the spectral weight underlying the Drude peak, is about 9000 cm$^{-1}$. As displayed in Fig. \ref{Fig2b}, the temperature dependence of the spectral weight $W=\int_0^{\Omega}\sigma_1(\omega)d\omega$, integrated up to $\Omega=\omega_p$, follows the quadratic relation $W(T)=W_0-BT^2$ \cite{ortolani05, toschi05}. The $T^2$ dependence of $W(T)$ is a general feature of metallic systems, that finds a natural explanation within a tight-binding approach, through the temperature dependence of the kinetic energy. In conventional metals, the energy scale set by $\omega_p$ should be large enough to fully recover the conservation of the spectral weight, resulting in a vanishingly small $B$ \cite{ortolani05}. 
The parameter $b=B/W_0$ gauges the presence of correlation effects \cite{baldassarre08}, which extend the temperature dependence of the spectral weight to energy scales larger than $\omega_p$. For LNO $b=3.2\times$10$^{-7}$K$^{-2}$, a value which is sizably larger than that found in gold (1.3$\times$10$^{-8}$K$^{-2}$) \cite{ortolani05}, and comparable to La$_{2-x}$Sr$_2$CuO$_4$ (2.5$\times$10$^{-7}$K$^{-2}$) \cite{ortolani05}. An alternative method to estimate correlations is to calculate the ratio of kinetic energies obtained from experiment and from band theory $K_{exp}/K_{band}$, which ranges from 0 for a Mott insulator, to 1 for conventional metals \cite{qazilbash09}. $K_{exp}$ is proportional to the spectral weight due to the Drude term. By integrating the optical conductivity up to 500 cm$^{-1}$ (which roughly corresponds to a cut-off for the Drude's spectral weight \cite{ouellette10}) and employing the LDA (Local Density Approximation) data from Ref. \cite{ouellette10} we find $K_{exp}/K_{band}\sim0.2$ for LNO, similar to underdoped and optimally doped cuprates \cite{qazilbash09}. 

The optical conductivity of the $n$=5 SL presents many analogies with pure LNO. Since a Drude peak is also well distinguishable in the data, we can again estimate the plasma frequency to be about 8500 cm$^{-1}$, roughly the same as for LNO. A $T^2$ behavior of $W(T)$ is found at temperatures between 150 and 400 K, with $b=1.6\times$10$^{-7}$K$^{-2}$. Below 150 K, which notably corresponds to the magnetic ordering temperature of bulk LMO and LNO/LMO SLs \cite{hoffman13}, a clear change is observed in the $T^2$ slope of $W(T)$, thus suggesting a coupling between magnetism and carrier density. The overall spectral weight is significantly larger in the $n$=5 compound, due to a larger mid-infrared conductivity.

In the compounds with $n<$5, the separation of the Drude term from the incoherent electronic background is more arbitrary. We can roughly estimate $\omega_p$ of about 5000 and 4000 cm$^{-1}$ for the $n$=4 and $n$=3 compounds respectively. $W(T)$ does not follow a quadratic law, and even increases with increasing temperature (panels a and b in Fig. \ref{Fig2b}), thus highlighting the increased role of incoherent excitations in shaping the infrared conductivity.
 
\begin{figure}
\includegraphics[width=9cm]{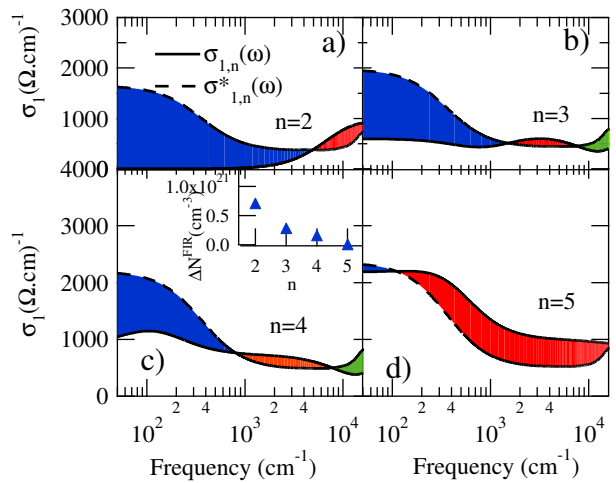}
\caption{Comparison between the room temperature $\sigma_{1,n}$, and $\sigma^*_{1,n}$ (see text) for $n$=2,3,4,5. The blue (red) area corresponds to spectral weight lost (gained) by the SL in the far- (mid-) infrared range due to the presence of interfacial effects. The green area highlights the transfer of spectral weight from higher energies.}
\label{Fig3}
\end{figure}

In order to better understand the present phenomenology, we introduce the average optical conductivity $\sigma^*_1$, defined as a weighted linear combination of the optical conductivity of the single constituents of a given SL:
\begin{equation}
\sigma^*_{1,n}=\frac{n\cdot\sigma_{1,LNO}+2\cdot\sigma_{1,LMO}}{n+2}.
\end{equation}
Since $\sigma^*_1$ disregards interfacial effects, the comparison between the measured $\sigma_{1,n}$ and $\sigma^*_{1,n}$ allows us to single out the features directly induced by the presence of the interfaces (Fig. \ref{Fig3}).

The presence of the interface induces a loss of  spectral weight $\Delta SW$ in the low frequency side, for all SL compounds (blue area in Fig. \ref{Fig3}). From this quantity, we can calculate the charge density ($\Delta N^{FIR}$) involved in such spectral weight redistribution by assuming $\Delta SW=\sqrt{\frac{4\pi \Delta N^{FIR}e^2}{m}}$, where $m=m_e$. As shown in the inset of Fig. \ref{Fig3}c, $\Delta N^{FIR}$ increases monotonically with decreasing $n$, i.e., by increasing the number of interfaces in the SL. 

The interfaces are also responsible for the piling up of new spectral weight at mid-infrared frequencies (red area in Fig. \ref{Fig3}). The mid-infrared spectral weight redistribution can not be systematically estimated for all compounds, since for $n$=2 and $n$=5, the red area extends above our maximum frequency limit (16000 cm$^{-1}$). However, we can safely conclude that the red area is always larger than the blue one  (note log scale on x axis in Fig. \ref{Fig3}). This indicates that the additional mid-infrared spectral weight is only in part due to the loss of spectral weight at low frequencies, and that a redistribution from the high frequency side (green area in Fig. \ref{Fig3}) is also at play.

\begin{figure}
\includegraphics[width=9cm]{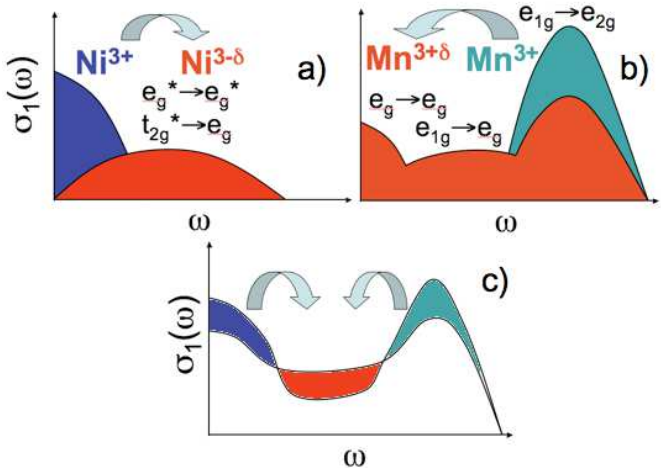}
\caption{Schematics of the spectral weight redistribution in LNO/LMO. a) Transfer of spectral weight from coherent (blue) to in-coherent (red) excitations in LNO, due to the reduction in the Ni oxidation state. b) Spectral weight transfer in LMO from the high energy Jahn-Teller band (green) to the mid and far infrared (red), due to the increased Mn valence. c) Sketch of the spectral weight transfer in LNO/LMO SLs, due to the valence mixing associated with interfacial effects.}
\label{Fig4}
\end{figure}

The optical data can be understood within the qualitative model depicted in Fig. \ref{Fig4}. In the superlattice compound, with decreasing $n$,  the valence of Ni decreases from its nominal Ni$^{3+}$ value typical of stoichiometric LaNiO$_3$. The reduction in the Ni oxidation state is known \cite{sanchez96} to drive a metal to insulator transition already at values about +2.75. From transport measurements on LNO reduced in oxygen \cite{sanchez96}, it is known that the semiconducting behavior is characterized by infrared activation energies (250 cm$^{-1}$ for Ni$^{2.75}$, and 1000 cm$^{-1}$ for Ni$^{2.5}$). We thus expect that the introduction of LMO layers in the SL, by reducing Ni valence, progressively depletes the low energy, coherent spectral weight of LNO, which piles up in the infrared, as schematically shown in Fig. \ref{Fig4}a. 

On the other hand, LMO is an insulator, with the first optical absorption occurring at about 2.5 eV \cite{okimoto97}, corresponding to the Jahn-Teller splitting of the $e_g$ band. When the Mn valence increases from 3+ to a higher oxidation state (as in La$_{1-x}$(Sr/Ca)$_x$MnO$_3$), polaronic states appear in the mid-infrared, whose spectral weight is taken from the Jahn-Teller band at 2.5 eV (see Fig. \ref{Fig4}b). 
At intermediate values of the Mn oxidation state  ($x=0.3\div0.5$) a metallic state may also set in, due to double-exchange physics \cite{zener51,millis95}. While we know from x-ray spectroscopy that Mn has mainly 4+ character in the SLs, the spectral weight increase observed below the magnetic ordering temperature in the inset of Fig. \ref{Fig2}d, may be a hint at the survival of the double-exchange physics in the LNO/LMO series.
Panel \ref{Fig4}c, pictorially describes the effect of the charge transfer between Mn and Ni sites on the overall optical conductivity, thus providing a qualitative explanation for the piling up of spectral weight at mid-infrared frequencies, at the expense of the far-infrared, and high energy (visible/UV) sides. Interfacial charge redistribution was previously identified as a possible origin of changes to the mid-infrared spectral response in manganite \cite{perucchi,choi11} and vanadate superlattices \cite{jeong11}. This is in contrast to what is observed in LNO/LaAlO$_3$ superlattices \cite{liu12,benckiser11} and in ultrathin LNO films \cite{scherwitzl11}, where localization is believed to occur due to dimensional confinement and enhanced correlations.

Infrared data show that the LNO/LMO superlattices display the presence of significant mid-infrared excitations that are not present in LNO or LMO alone. Since SLs are intrinsically ordered structures, we can rule out disorder as a possible source for the presence of such incoherent excitations. On the other hand, both electronic correlation and polaronic effects (or even more likely, a mixture of the two) can be invoked as the mechanisms underlying the piling up of spectral weight in the mid infrared range. By increasing $n$, it is possible to progressively enhance the coherent excitations, thereby tuning the degree of metallicity of the SL. For $n$=5 the low energy electrodynamics is dominated by coherent excitations, while some coupling with magnetic ordering takes place below 150 K. Ab initio calculations could be of help in better elucidating these points \cite{baldassarre08}. The LNO/LMO systems thus provide an exciting new platform to manipulate and control the interplay of electronic, magnetic and vibrational degrees of freedom in a disorder-free 2D oxide material. 

\begin{acknowledgments}

This work was partially supported by Italian Ministry of Research (MIUR) program FIRB Futuro in Ricerca grant no. RBFR10PSK4. J.D.H and A.B. acknowledge support from Department of Energy, Office of Basic Energy Science, Materials Science and Engineering Division. Work at Argonne National Laboratory, including the use of the Center for Nanoscale Materials and Advanced Photon Source, was supported by the U.S. Department of Energy, Office of Basic Energy Sciences under contract number DE-AC02-06CH11357.

\end{acknowledgments}

\end{document}